\documentclass[fleqn,usenatbib]{mnras}
\usepackage{newtxtext,newtxmath}
\usepackage[T1]{fontenc}
\usepackage{ae,aecompl} % for Overleaf
\usepackage{graphicx} % Including figure files
\usepackage{amsmath}  % Advanced maths commands
\usepackage{amssymb}  % Extra maths symbols

\newcommand{\Mpi}{\piup}
\newcommand{\msun}{{\rm M}_\odot}

\newcommand{\nh}{n_{\rm H}}

\newcommand{\cc}{{\rm cm^{-3}}}
\newcommand{\kms}{{\rm km\,s^{-1}}}
\newcommand{\mDM}{m_{\rm DM}}
\newcommand{\cmcm}{{\rm cm^{2}}}
\newcommand{\vbDM}{v_{\rm bDM}}
\newcommand{\Tcmb}{T_{\rm CMB}}
\newcommand{\Tvir}{T_{\rm vir}}

\title
[First stars with baryon-DM scattering]%[37/45 characters]
{Baryon-dark matter scattering and first star formation}

\author[S. Hirano \& V. Bromm]{
Shingo Hirano,$^{1}$\thanks{E-mail: hirano0613@gmail.com}
and
Volker Bromm,$^{1}$
\\
$^{1}$Department of Astronomy, University of Texas at Austin, Austin, TX 78712, USA
}

\date{Accepted XXX. Received YYY; in original form ZZZ}
\pubyear{2018}

\begin{document}
\label{firstpage}
\pagerange{\pageref{firstpage}--\pageref{lastpage}}
\maketitle

\begin{abstract}%[198/200 words for Letters]
The recent detection of the sky-averaged 21-cm cosmological signal indicates a stronger absorption than the maximum allowed value based on the standard model. One explanation for the required colder primordial gas is the energy transfer between the baryon and dark matter fluids due to non-gravitational scattering. Here, we explore the thermal evolution of primordial gas, collapsing to form Population~III (Pop~III) stars, when this energy transfer is included. Performing a series of one-zone calculations, we find that the evolution results in stars more massive than in the standard model, provided that the dark matter is described by the best-fit parameters inferred from the 21-cm observation. On the other hand, a significant part of the dark matter parameter space can be excluded by the requirement to form massive Pop~III stars sufficiently early in cosmic history. Otherwise, the radiation background needed to bring about the strong Wouthuysen-Field coupling at $z\gtrsim17$, inferred to explain the 21-cm absorption feature, could not be builtup. Intriguingly, the independent constraint from the physics of first star formation at high densities points to a similarly narrow range in dark matter properties. This exploratory study has to be followed-up with self-consistent three-dimensional simulations for a more rigorous derivation.
\end{abstract}


%%%%%%%%%%%%%%%%%%%%%%%%%%%%%%%%%%%%%%%%%%%%%%%%%%
%%%%%%%%%%%%%%%%% BODY OF PAPER %%%%%%%%%%%%

%%%%%%%%%%%%%%%%%%%%%%%%%%%%%%%%%%%%%%%%%%%%%%%%%%
%%%%% 1.
\section{Introduction}
\label{sec:intro}

%[1-1]
One of the fundamental open questions in modern science is to elucidate the nature of dark matter (DM).
Due to the lack of any direct detections in the laboratory yet, astrophysical probes have gained increasing prominence. 
Specifically, the detailed properties of the DM particle(s) are reflected in cosmological structure formation, with a particular sensitivity on small-scales. 
This regime is accessible either locally by probing dwarf galaxies in the Local Group \citep[e.g.][]{bullock17}, or at high redshifts, when the first stars form in low-mass DM haloes \citep[e.g.][]{fialkov12,bromm13,dayal17a}. 
This latter probe will be rendered even more powerful when the {\it James Webb Space Telescope (JWST)} and other frontier facilities will become available in the near future \citep[e.g.][]{dayal17b}.

%[1-2]
An alternative pathway to first star formation is given by 21-cm cosmology, which probes the redshifted radiation emitted in the spin-flip transition of neutral hydrogen \citep[reviewed in][]{furlanetto06}. 
The Experiment to Detect the Global Epoch of Reionization Signature (EDGES) has recently reported the spectral absorption feature when the spin temperature is coupled to that of the cold intergalactic medium (IGM) gas \citep{bowman18nat}. 
This coupling is mediated through the Lyman-$\alpha$ radiation produced by the first stars, the so-called Wouthuysen-Field effect \citep{wouthuysen52,field58}.
The best-fit absorption profile is centred at a frequency of $78 \pm 1$\,MHz, corresponding to $z = 17.2$, with a brightness temperature of $T_{21} = -500^{+200}_{-500}$\,mK. 
This represents an unexpectedly deep absorption, compared to the expectation within the standard model \citep[][]{pritchard12}.
To explain the strength of the absorption signal, the primordial gas must have been colder than expected. 
Any astrophysical phenomena could only have acted to raise the IGM gas temperature, such as the heating from early sources of X-ray radiation \citep[e.g.][]{jeon12}, and thus cannot explain it.

%[1-3]
One of the theoretical explanations for the additional cooling is non-standard DM particle physics in the form of baryon-dark matter (b-DM) scattering \citep{barkana18nat}.
Such non-gravitational interaction may act to solve a number of discrepancies between the standard cold dark matter (CDM) cosmological model and observational constraints \citep[e.g.][]{deblok10,boylan-kolchin12,bullock13}.
Because the DM fluid decouples from the cosmic microwave background (CMB) earlier than the baryons, the latter can be cooled by the colder DM component via b-DM scattering. 
The baryons, thus cooled, can imprint their signature on the redshifted 21-cm line from the neutral hydrogen during cosmic dawn when the Wouthuysen-Field effect re-established the spin coupling to the gas temperature \citep{tashiro14}.
Additionally, the baryons could also thermalize with a kinematically hotter DM component \citep{munoz15}, such as the supersonic streaming velocity left over after cosmic recombination \citep{tseliakhovich10}.

%[1-4]
In the case with b-DM scattering, other astrophysical phenomena can be affected, in addition to the impact on the 21-cm signal, such as the properties of the Lyman-$\alpha$ forest \citep{dvorkin14} and the abundance of high-$z$ galaxies \citep[e.g.][]{mirocha18}. 
All of these effects, however, probe the low-density diffuse regime, and it is an open question how such potential b-DM scattering would influence the collapse of primordial gas to the high densities where star formation occurs. 
It is well known that the formation physics of the first, so-called Population~III (Pop~III), stars can depend on the particle-physics nature of DM \citep[e.g.][]{stacy14,hirano18fdm}.
We here specifically ask: Does a model with b-DM scattering, calibrated at low densities, result in Pop~III star formation at high densities that is consistent with the low-density assumptions? 
To achieve the early Wouthuysen-Field coupling implied by the EDGES measurement, Pop~III stars would need to produce a critical background Lyman-$\alpha$ intensity of $1.8 \times 10^{-21} [(1+z)/20]\,{\rm erg\,s^{-1}\,cm^{-2}\,Hz^{-1}\,st^{-1}}$\citep{madau97,ciardi2003}.
The ionizing stellar UV radiation is reprocessed into Lyman-$\alpha$ photons in the surrounding emission nebula, and the ionizing rate per stellar mass is higher for more massive stars \citep{bromm01,schaerer02}. We estimate that Pop~III stars characterized by a top-heavy initial mass function (IMF) are required to provide the critical intensity for effective Wouthuysen-Field coupling.
How then do the first stars form in a model with b-DM scattering?

%[1-5]
We specifically wish to assess how important this effect is for primordial stars in minihaloes, the standard formation site in $\Lambda$CDM cosmology.
Ultimately, self-consistent cosmological simulations including the b-DM scattering throughout are necessary to reach definitive conclusions.  
Here, we carry out a first survey of the effect, and evaluate the energy transfer rate between the baryon and DM components with a series of one-zone calculations, thus assessing the dependence on model parameters.

%%%%%%%%%%%%%%%%%%%%%%%%%%%%%%%%%%%%%%%%%%%%%%%%%%
%%%%% 2.
\section{Modified thermal evolution}
\label{sec:model}

%[2-1]
As the most important impact of b-DM scattering on the star-formation process, we consider the energy transfer between the baryon and DM components.
We assume a scattering cross section, $\sigma(\vbDM) = \sigma_1\,(\vbDM/1\,\kms)^{-4}$, as a function of the b-DM relative velocity $\vbDM$.
This general dependence is often considered, and would formally correspond to a Coulomb-like b-DM scattering \citep[e.g.][]{tashiro14}.
The energy transfer rate, $\dot{Q}_{\rm bDM}$, is calculated with equation~(16) in \cite{munoz15},
\begin{eqnarray}
	\dot{Q}_{\rm bDM} &=& \frac{m_{\rm b} \rho_{\rm DM} \sigma_1}{(\mDM + m_{\rm b})^2} \Biggl[ \frac{(T_{\rm DM}-T_{\rm b})}{u_{\rm th}^3} \left\{ \sqrt{\frac{2}{\Mpi}} e^{-\frac{r^2}{2}} \right\} \nonumber \\
    &&+ \frac{\mDM}{\vbDM} \left\{ {\rm erf} \left( \frac{r}{\sqrt{2}} \right) - \sqrt{\frac{2}{\Mpi}} e^{-\frac{r^2}{2}} r \right\} \Biggr] \,,
\label{eq:dQdt}
\end{eqnarray}
where $m_{\rm b}$ and $\mDM$ are proton and DM particle mass, $\rho_{\rm DM}$ is the density of the DM fluid, $T_{\rm b}$ and $T_{\rm DM}$ (where $(1/2) \mDM v^2 = (3/2) k_{\rm B} T_{\rm DM}$) are (effective) temperatures of the two fluids, $u_{\rm th} \equiv \sqrt{T_{\rm b}/m_{\rm b} + T_{\rm DM}/\mDM}$ is the variance of the thermal relative motion of the two components, and $r \equiv \vbDM/u_{\rm th}$ indicates which term is more important.
The first term, giving the temperature-dependent heating (cooling), dominates when $r \ll 1$ ($\vbDM \ll u_{\rm th}$).
The second term, on the other hand, describing heating due to the relative velocity, becomes important when $r \gg 1$ ($\vbDM \gg u_{\rm th}$), where the expression in the curly bracket increases from zero to one for $r = 0 \to \infty$. Based on the DM properties, the two key parameters are: $\mDM$ governs which term is dominant, and $\sigma_1$ determines the strength of the total contribution.
Here, we adopt the best-fit parameter set from the recent 21-cm observation, ($\mDM$/GeV, $\sigma_1/\cmcm$) = ($0.3$, $8 \times 10^{-20}$), as fiducial values for these parameters \citep{barkana18nat}.

%----------------------------------------
\begin{figure}
\includegraphics[width=0.9\columnwidth]{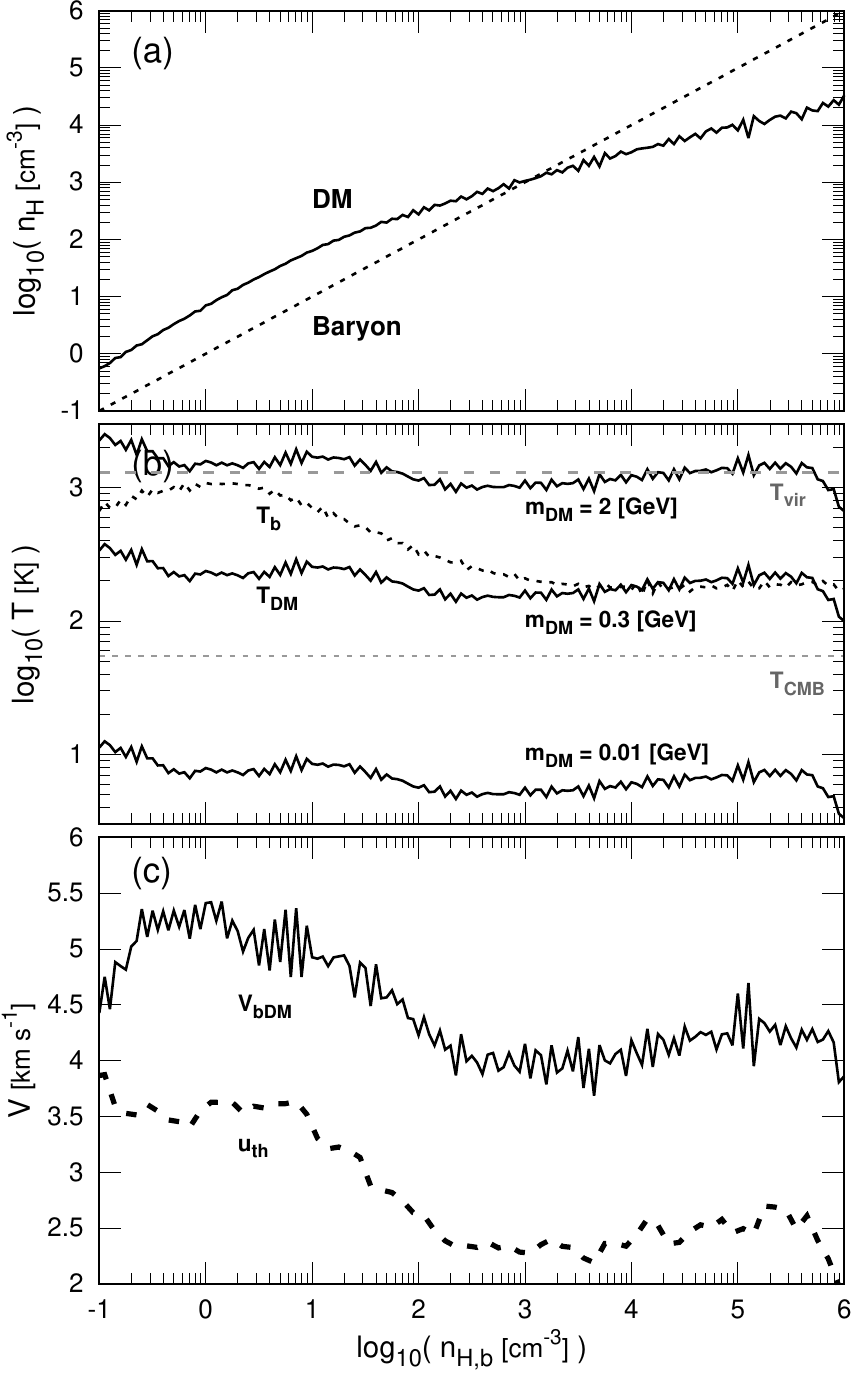}
\caption{
Profiles of select physical properties for primordial gas clouds collapsing into minihaloes: 
(a) DM and baryon number densities, normalized to the proton mass; 
(b) gas temperature ({\it dotted line}) and effective DM temperature ({\it solid lines}); and 
(c) relative velocity between the two fluids ($\vbDM$) and variance of the thermal motion ($u_{\rm th}$).
We average over 110 minihaloes in a large-scale cosmological simulation box, when the collapsing gas reaches $\nh = 10^6\,\cc$ \citep{hirano14}.
The three DM temperatures in panel (b) are for $\mDM/{\rm GeV} = 2$, $0.3$, and $0.01$, respectively.
The horizontal lines in panel (b) represent the virial and CMB temperatures for a typical minihalo of $M_{\rm vir} = 3 \times 10^5\,\msun$ at $z = 19$: $\Tvir = 1.3 \times 10^3$\,K and $\Tcmb = 55$\,K.
}
\label{fig1}
\end{figure}
%----------------------------------------

%[2-2]
To compute $\dot{Q}_{\rm bDM}$ in the primordial star-forming gas cloud, we derive the typical properties in a host minihalo from three-dimensional cosmological hydrodynamic simulations \citep{hirano14}.
The simulations were carried out in the standard $\Lambda$CDM cosmology without b-DM scattering. 
We thus make the idealizing assumption that the early stages of structure formation, occurring at low densities, is not significantly different across models that include the b-DM scattering effect. 
Evidently, this needs to be tested in future, self-consistent cosmological simulations, but we present justification for this approximation below, at least for an important part of parameter space (see Section~\ref{sec:1zone}). 
Based on the cosmological simulations, we analyse averaged properties of $\rho_{\rm b}$, $\rho_{\rm DM}$, $T_{\rm b}$, $T_{\rm DM}$, and $\vbDM$.
Fig.~\ref{fig1} presents averaged profiles based on 110 primordial star-forming gas clouds.

%[2-3]
We use the snapshots when the central baryon number density reaches $10^6\,\cc$, so that we can resolve the collapse stage where the temperature assumes a minimum, and the cloud becomes gravitationally unstable, the so-called loitering point \citep{bromm02}. 
This scale, closely related to the pre-stellar core in present-day star formation \citep[e.g.][]{mckee07}, describes the overall mass scale of Pop~III stars (see below).
During the cloud collapse, the baryon density exceeds the DM density at $\nh \sim 10^3\,\cc$ ({\it panel a}).
For our fiducial parameters, the DM effective temperature
%, $\mDM v^2 \sim k_{\rm B} T_{\rm DM}$, 
remains below the gas temperature until the loitering scale, $\nh < 10^3\,\cc$, and becomes comparable during the subsequent collapse ({\it panel b}). 
Thus, the first term of $\dot{Q}_{\rm bDM}$ is a cooling term in the early phase, but acts as a heating term later on. 
The relative velocity between the two fluids, calculated as the mean-square average inside the smoothed particle hydrodynamics (SPH) spline kernel, $\vbDM = \sqrt{ (1/N) \sum^N_{i=0} (\textbf{\textit{v}}_{\rm b} - \textbf{\textit{v}}_{\rm DM,i})^2 }$, is about $4-5\,\kms$. 
This is larger than the typical velocities discussed in the literature to explain the 21-cm observation ($\sim\!1\,\kms$), but is of order the virial velocity in a typical Pop~III host minihalo.
The heating from the second term becomes more important in the star-forming cloud during the later collapse stage.

%----------------------------------------
\begin{figure}
\includegraphics[width=0.9\columnwidth]{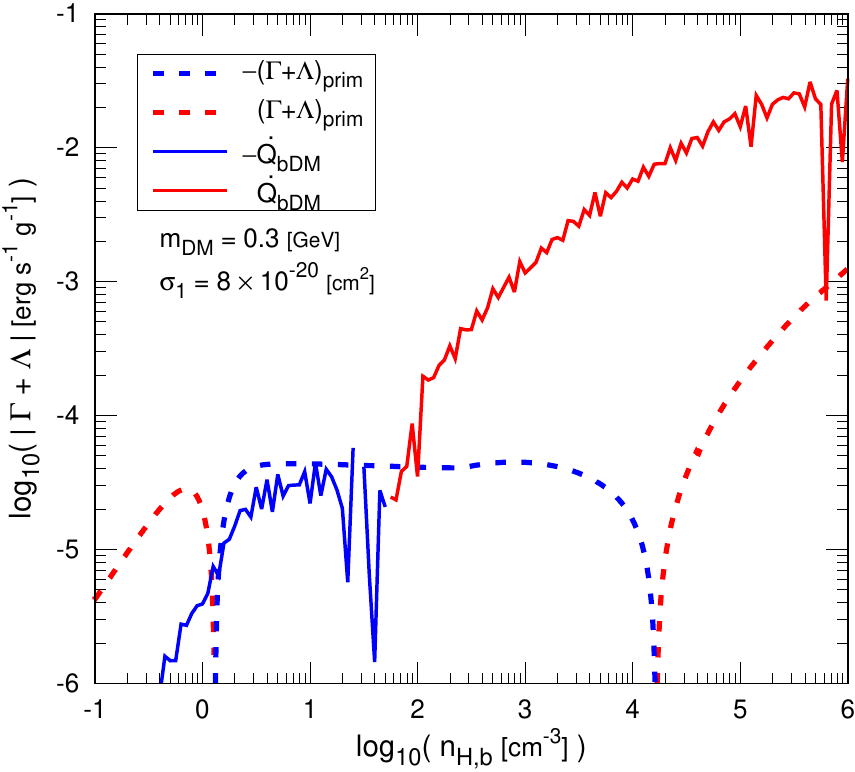}
\caption{
Energy transfer rate $\dot{Q}_{\rm bDM}$ (equation~\ref{eq:dQdt}), with the fiducial parameter set of ($\mDM$/GeV, $\sigma_1/\cmcm$) = ($0.3$, $8 \times 10^{-20}$), as a function of baryon number density.
The dashed lines represent the net energy transfer rate for the collapsing primordial gas cloud in the standard case without the b-DM scattering. 
Here, the early thermal evolution, up to densities of $\sim\!100\,\cc$, is the same as in the standard collapse case, whereas the heating from the b-DM interaction dominates towards later collapse stages.
}
\label{fig2}
\end{figure}
%----------------------------------------

%[2-4]
Fig.~\ref{fig2} shows the energy transfer rate $\dot{Q}_{\rm bDM}$ for the representative primordial star-forming cloud (Fig.~\ref{fig1}) with the fiducial parameters.
The energy transfer behaviour switches from cooling to heating when the gas cloud contracts inside the host DM minihalo beyond $\nh \sim 100\,\cc$.
During the initial collapse phase, the possible thermal impact from b-DM scattering is weaker than that from the standard primordial chemistry and cooling ($|\dot{Q}_{\rm bDM}| < |\Gamma + \Lambda|_{\rm prim}$).
Once the collapse has proceeded to higher densities, however, the heating due to b-DM scattering overtakes the standard contribution ($|\dot{Q}_{\rm bDM}| > |\Gamma + \Lambda|_{\rm prim}$). This heating and cooling sequence will influence the thermodynamics of primordial star-forming gas, modifying the mass scale for gravitational instability. As a consequence, the typical mass of the first stars will also be affected, as we will discuss next.

%%%%%%%%%%%%%%%%%%%%%%%%%%%%%%%%%%%%%%%%%%%%%%%%%%
%%%%% 3.
\section{Collapse and fragmentation}
\label{sec:1zone}

\begin{figure}
\includegraphics[width=0.9\columnwidth]{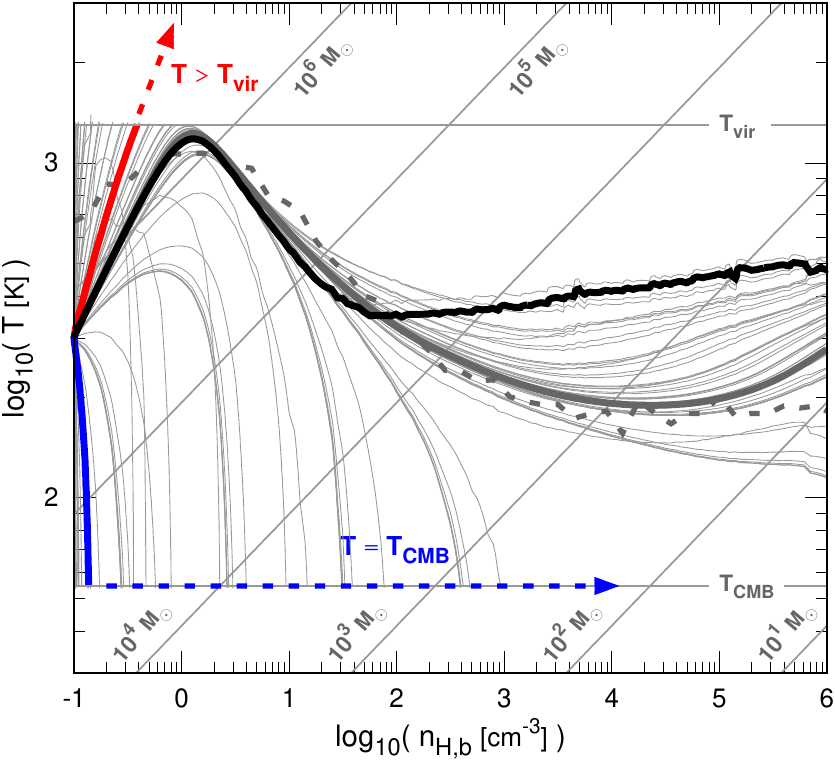}
\caption{
Thermal evolution of the collapsing primordial gas cloud evaluated with idealized one-zone calculations.
The red, black, and blue lines show results for select parameters, as follows: ($\mDM$/GeV, $\sigma_1/\cmcm$) = (2, $3 \times 10^{-19}$), (0.3, $8 \times 10^{-20}$), and (0.01, $1 \times 10^{-18}$), respectively. 
Thin lines show cases for additional parameter choices.
The thick grey line represents the standard case without b-DM energy transfer, and well reproduces the averaged data for an ensemble of cosmological clouds ({\it dashed line}).
The diagonal lines show the $\nh-T$ relation for the given BE masses.
The extreme cases, of either driving the gas temperature above the minihalo virial temperature ($T > T_{\rm vir}$; {\it red line}), or quickly cooling to the CMB floor ($T = T_{\rm CMB}$; {\it blue}), are not well captured by our simplified modelling.
The long-term thermal evolution is therefore uncertain, as indicated by the dashed arrows.  
}
\label{fig3}
\end{figure}

%[3-1]
To assess the fragmentation properties of primordial gas, we introduce the energy transfer rate due to the b-DM scattering to an one-zone calculation to model the thermal evolution of a cloud collapsing into a minihalo. 
We employ the same chemical network as in the cosmological simulation used to derive the typical minihalo properties \citep{hirano14}.
The thermal evolution of the collapsing cloud depends on the collapse rate \citep[e.g.][]{chiaki16}, and we adjust the parameters of our one-zone model to reproduce the averaged data from the cosmological simulation.
In Fig.~\ref{fig3}, we show three different thermal pathways due to the additional energy transfer rate, depending on the nature of the DM.

%----------------------------------------
%----------------------------------------

%[3-2]
${\bf T_{\rm CMB} < T < T_{\rm vir}:}$
The solid black line in Fig.~\ref{fig3} shows the thermal evolution under the influence of b-DM scattering, with the fiducial parameters which can well explain the deep absorption in the detected 21-cm spectrum. 
The resulting thermal history is almost the same as for the standard case without b-DM interaction until $\nh \sim 100\,\cc$, because of the negligible contribution from the additional cooling and heating terms in this density regime (see Fig.~\ref{fig2}). 
We note that this supports our idealizing assumption that the standard $\Lambda$CDM cosmological simulations provide a valid representation of the initial stages of structure formation, even if non-gravitational b-DM scattering were present.
Towards higher densities, however, the collapsing cloud heats up strongly due to the energy exchange from the b-DM scattering. 
Assuming that the gas cloud becomes gravitationally unstable when the temperature reaches its minimum value during the collapse, the fragmentation scale can be estimated via the Bonnor-Ebert (BE) mass, $M_{\rm BE} \approx 1000\,\msun\,(T/200\,{\rm K})^{3/2}\,(n_{\rm H}/10^4\,\cc)^{-1/2}$ \citep[][]{abel02,bromm02}. 
The BE mass in the fiducial b-DM scattering case (the black line in Fig.~\ref{fig3}) is then $\sim 2 \times 10^4\,\msun$, about 40 times larger than the case without such scattering (the thick grey line), where $M_{\rm BE}\sim 500\,\msun$.

%[3-3]
Because the mass accretion rate onto a protostar depends on the gas temperature, $\dot{M} = M_{\rm BE} / t_{\rm ff} \propto T^{3/2}$, such hot cloud core could host more massive Pop~III stars. 
The conclusion is that very massive first stars could form in a cosmology with non-gravitational b-DM scattering, if described by the 21-cm best-fit parameters. 
This also demonstrates that, in this case, the low-density physics, explaining the deep 21-cm absorption feature, is consistent with the star formation physics in the high-density regime. 
The latter indicates that Pop~III stars can be sufficiently massive to produce the high-energy radiation needed to mediate the Wouthuysen-Field coupling at early cosmic epochs. 
We also perform a series of additional one-zone calculations to cover parameter space, $10^{-4} \le \mDM/{\rm GeV} \le 10^{2}$ and $10^{-21} \le \sigma_1/\cmcm \le 10^{-16}$, the same range shown in fig.~3 of \cite{barkana18nat}.
We show a subset of the cases with thin lines in Fig.~\ref{fig3} for select parameter choices, comparing them with the standard $\Lambda$CDM case without b-DM scattering.
In addition, there are two distinct thermal evolution modes, due to strong cooling or heating from the b-DM interaction, that result in situations where the formation of massive Pop~III stars is disfavoured.

%[3-4]
${\bf T \simeq T_{\rm CMB}:}$
If the cooling by the DM fluid is too strong ($T_{\rm DM} \ll T_{\rm b}$ when $\mDM \ll m_{\rm b}$), the gas temperature is tied to the CMB temperature floor, $\Tcmb = 2.73\,{\rm K}\,(1+z) \simeq 55\,{\rm K}\,(1+z)/20$, assuming a typical formation redshift of $z \simeq 20$.
After the CMB floor is reached, the cloud's evolution becomes nearly isothermal, and the final fragmentation mass scale is $M_{\rm BE} \le 10\,\msun$. 
In extreme cases, primordial gas could cool even before the DM minihalo gravitationally compresses the baryon fluid, resulting in a completely different pathway to star formation. 
In any case, the primordial gas cloud cooled to the CMB value fragments into small clumps, forming predominantly low-mass Pop~III stars, which less efficiently produce nebular Lyman-$\alpha$ flux per stellar mass.
This is inconsistent with the inferred Wouthuysen-Field coupling implied by the EDGES detection which requires copious ionizing stellar UV radiation from massive stars in the early Universe.

%[3-5]
${\bf T > T_{\rm vir}:}$
The opposite case with strong heating ($T_{\rm DM} \gg T_{\rm b}$ when $\mDM \gg m_{\rm b}$) would also be excluded, because of the difficulty to explain the observed strong 21-cm absorption signal at $z = 17.2$. 
If the effective (kinetic) DM temperature exceeded the virial temperature of the minihalo, the gas cloud could not begin to collapse until the halo had sufficiently grown beyond its fiducial value\footnote{As parameters of the host minihalo, we adopt the average value obtained from a large cosmological sample \citep[see fig.~3 of][]{hirano15}: $z = 19$ and $M_{\rm h} = 3 \times 10^5\,\msun$.}, resulting in an increased virial temperature, $\Tvir \simeq 1.3 \times 10^3\,{\rm K}\,(M_{\rm h}/3 \times 10^5\,\msun)^{2/3}\,(1+z)/20$ \citep{barkana01,bromm13}.
We terminate the one-zone calculation when $T$ reaches $T_{\rm vir}$, because the properties of the baryon fluid were pre-calculated from data of clouds formed in minihaloes. 
In the case of DM with such parameter sets, the formation of Pop~III stars would be delayed until more massive haloes finally emerge, inconsistent with the detection of the 21-cm absorption trough already at $z \simeq 17$, which indicates the presence of radiation sources early on.

%%%%%%%%%%%%%%%%%%%%%%%%%%%%%%%%%%%%%%%%%%%%%%%%%%
%%%%% 4.
\section{Constraints on DM physics}
\label{sec:dis}

%----------------------------------------
\begin{figure}
\includegraphics[width=1.0\columnwidth]{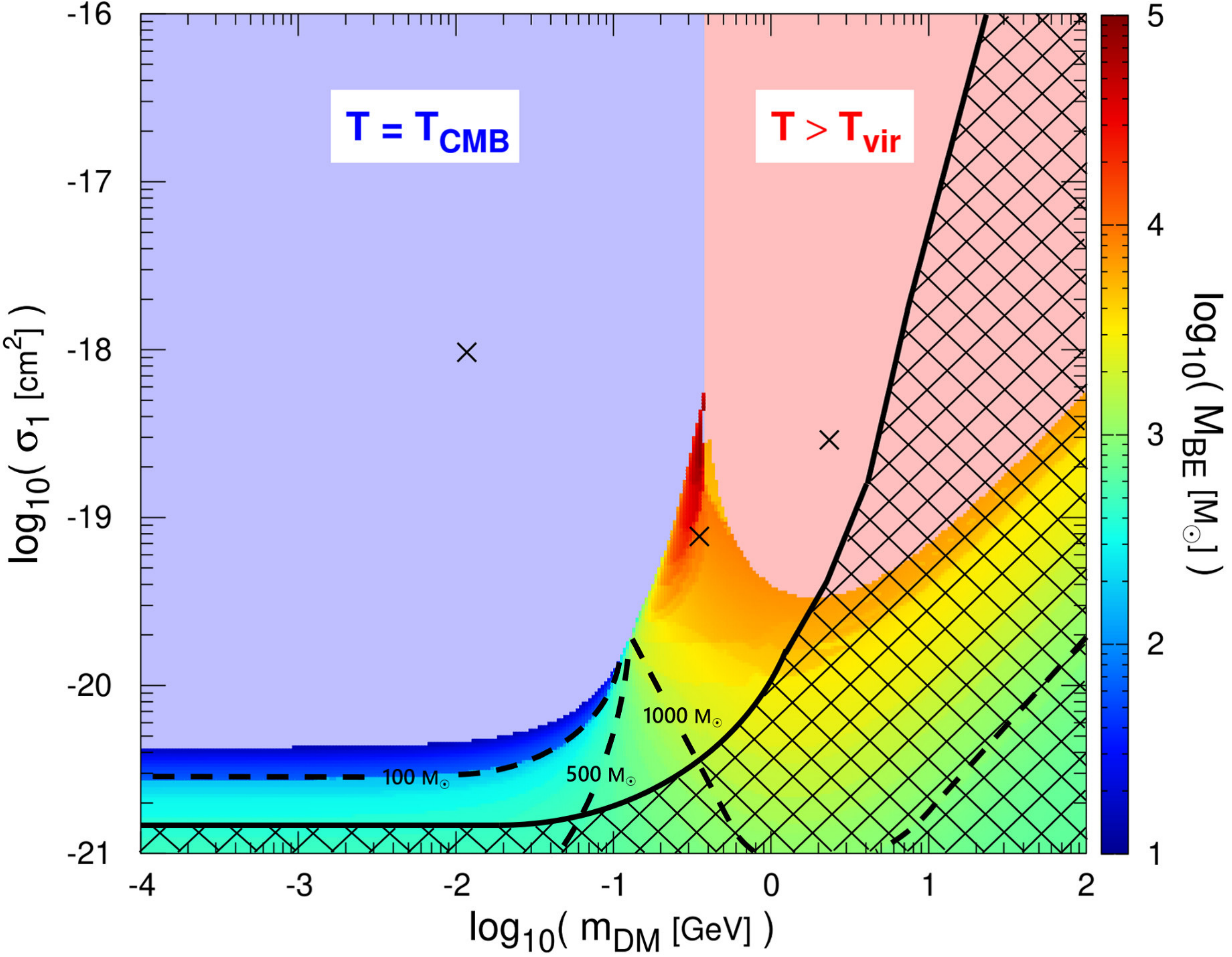}
\caption{
Constraints on DM properties, $\mDM$ and $\sigma_1$, using first star formation. 
The physics of primordial star formation delineates regions where the formation of massive stars at early epochs is disfavoured: 
(1) the gas temperature reaches the CMB floor, thus leading to predominantly low-mass stars ({\it light-blue region}) and 
(2) the gas temperature exceeds the minihalo's virial temperature, thus delaying the collapse and onset of star formation ({\it light-red region}). 
For the region where massive stars can form, on the other hand, we indicate the Bonner-Ebert mass with the coloured contours.
The three dashed lines show the $\mDM-\sigma_1$ relation for the given BE masses, specifically $M_{\rm BE} = 100$, $500$, and $1000\,\msun$, respectively.
Finally, the hatched region shows the region excluded by EDGES, assuming absorption by at least $T_{21} = -231$\,mK at $z = 17$ \citep[see fig.~3 in][]{barkana18nat}.
The three crosses correspond to the three parameter sets used in Figs.~\ref{fig1} and \ref{fig3}. 
As can be seen, there is only a narrow parameter range where all available constraints are fulfilled.
}
\label{fig4}
\end{figure}
%----------------------------------------

%[4-1]
In Fig.~\ref{fig4}, we summarize the calculation results, in terms of the dependence on DM parameter space, $\mDM$ and $\sigma_1$. 
The b-DM scattering acts as an additional cooling source for low $\mDM$, whereas as heating for large $\mDM$. 
The overall thermal impact increases with $\sigma_1$. 
The colour contour represents the mass scale for gravitational instability, $M_{\rm BE}$, whereas the two light-coloured zones indicate regions in parameter space that are excluded, because of constraints on Pop~III star formation, as discussed in the previous section. 
Furthermore, any region is effectively excluded, where $M_{\rm BE}$ is too low to accommodate a top-heavy IMF, as is the case for low values of $\mDM$. 
The hatched region reproduces the constraints from the EDGES measurement, assuming the 3.5$\sigma$ observational results: $T_{21} = -231 $\,mK at $z = 17$ \citep{barkana18nat}. 
Intriguingly, the allowed parameter range is highly confined by combining constraints derived from high- and low-density physics.

%[4-2]
The case with the 21-cm observation best-fit DM parameter set is located in the region where the energy transfer due to b-DM scattering changes from cooling to heating during the cloud collapse (Fig.~\ref{fig2}).
Around $\mDM \sim 0.35$\,GeV, the b-DM cooling is negligible, or at most comparable to the cooling provided by the standard primordial chemistry, and cannot affect the collapse of the baryon fluid, gravitationally trapped inside a DM minihalo. 
Subsequently, however, the efficient b-DM heating can greatly change the thermal evolution of the collapsing gas cloud. 
As a result, the mass scale of a gravitationally unstable object significantly increases, to $M_{\rm BE} \sim\!10^4\,\msun$ compared with the standard $\Lambda$CDM case ($\sim\!1000\,\msun$).

%[4-3]
Our series of one-zone models is capable of exploring the basic DM parameter dependence of first star formation. 
In so doing, our calculations adopt a number of idealizing approximations which could affect the results. 
First, our methodology ignores the direct b-DM momentum transfer in the force equation, and the thermal back reaction on the DM fluid due to b-DM scattering, which could moderate the overall thermal impact on the gas. 
Consequently, our results represent the maximal effect on Pop~III star-formation.
Second, we derive the physical properties of primordial star-forming clouds from cosmological simulations that were performed without considering the b-DM scattering. 
Clearly, fully self-consistent cosmological simulations need to be carried out in future work to test the validity of our approach.

%[4-4]
The initial moments of star formation in the Universe, marking the end of the cosmic dark ages, provide a powerful probe of the particle physics nature of DM. 
Remarkably, if DM had the characteristics inferred from the recent 21-cm EDGES result, including the non-gravitational b-DM scattering, self-consistent Pop~III star formation would require a similar range in DM parameter space. 
This range, preferred both by 21-cm cosmology and by the star formation physics, would notably include a DM particle mass of order the proton mass. 
Such low masses, compared to the expectation for the generic supersymmetric weakly interacting massive particle (WIMP), would also help to understand the lack of any direct detections yet \citep[e.g.][]{bertone10}. 
Maybe the first stars may thus finally elucidate the most elusive matter component of the Universe.

\section*{Acknowledgements}

This work was supported by Grant-in-Aid for JSPS Overseas Research Fellowships to SH, and by NSF grant AST-1413501 to VB.

%%%%%%%%%%%%%%%%%%%%%%%%%%%%%%%%%%%%%%%%%%%%%%%%%%
%%%%% REFERENCES %%%%%%%%%%%%%%%%%%%%%%%%%%%%%%%%%

\bibliographystyle{mnras}
\bibliography{biblio}

%%%%%%%%%%%%%%%%%%%%%%%%%%%%%%%%%%%%%%%%%%%%%%%%%%
%%%%% Don't change these lines %%%%%%%%%%%%%%%%%%%
\bsp	% typesetting comment
\label{lastpage}
\end{document}